\pdfoutput=1

\documentclass[11pt]{article}

\usepackage{emnlp2021}

\usepackage{times}
\usepackage{latexsym}
\usepackage{xspace}

\usepackage[T1]{fontenc}

\usepackage[utf8]{inputenc}

\usepackage{microtype}

\usepackage{times}
\usepackage{latexsym}
\usepackage{enumitem}
\usepackage{paralist}
\usepackage{caption}
\usepackage{subcaption}

\usepackage{comment}
\usepackage[flushleft]{threeparttable}
\usepackage{graphicx}
\usepackage{amsmath}
\usepackage{tabularx}
\usepackage{mathrsfs}
\usepackage{multirow}
\usepackage{algorithm}
\usepackage{enumitem}
\usepackage{xcolor}
\usepackage{listings}
\usepackage{hyperref}
\usepackage{arydshln}

\usepackage{amssymb}
\usepackage{pifont}
\newcommand{\cmark}{\ding{51}}%
\newcommand{\xmark}{\ding{55}}%

\usepackage[utf8]{inputenc}
\usepackage{cleveref}
\crefname{section}{§}{§§}
\Crefname{section}{§}{§§}

\usepackage[frozencache,cachedir=.]{minted}

\usepackage{xcolor} 
\usepackage{tcolorbox}

\definecolor{blueish}{RGB}{250, 250, 255}
\definecolor{greenish}{RGB}{250, 255, 250}
\newtcbox{\inlinebox}[1][]{
 box align=base,
 nobeforeafter,
 colback=blueish,
 size=small,
 left=0pt,
 right=0pt,
 boxsep=2pt,
 #1}

\newcommand{\highlight}[1]{%
{%
\inlinebox{#1}%
}}

\definecolor{MyColor}{RGB}{50, 100, 250}
\definecolor{Orange}{RGB}{244, 101, 66}
\definecolor{Red}{RGB}{255, 0, 0}
\definecolor{Green}{RGB}{0, 255, 0}
\definecolor{Blue}{RGB}{0, 0, 255}
\newcommand{\notesc}[1]{\textcolor{purple}{\bf\small [#1 --Saikat]}}

\newcommand{\tool}{{REDCODER}\xspace}
\newcommand{\toolnospace}{REDCODER}
\newcommand{\toolext}{{REDCODER-EXT}\xspace}
\newcommand{\toolextnospace}{REDCODER-EXT}
\newcommand{\toolcode}{REDCODER-EXT\xspace}
\newcommand{\coder}{SCODE-R\xspace}

\newcommand{\eg}{\textit{e.g.,}~}
\newcommand{\ie}{\textit{i.e.,}~}
\newcommand{\wrt}{\textit{w.r.t.}~}
\newcommand{\aka}{\textit{a.k.a.}~}

\usepackage{arydshln}
\usepackage[belowskip=0pt,aboveskip=5pt]{caption}
\usepackage{fdsymbol}

\title{Retrieval Augmented Code Generation and Summarization}

\author{
Md Rizwan Parvez$^\S$, Wasi Uddin Ahmad$^\S$, Saikat Chakraborty$^\dagger$ \\
\textbf{Baishakhi Ray$^\dagger$, Kai-Wei Chang$^\S$} \\
$^\S$University of California, Los Angeles, $^\dagger$Columbia University \\
$^\S${\{rizwan, wasiahmad, kwchang\}@cs.ucla.edu}, $^\dagger${\{saikatc, rayb\}@cs.columbia.edu}
}

\lstdefinestyle{CustomPy}{
  belowcaptionskip=1\baselineskip,
  xleftmargin=2pt,
  xrightmargin=1pt,
  language=Python,
  numbersep=5pt,
  captionpos=b,
  tabsize=3,
  showstringspaces=false,
  basicstyle=\fontsize{9}{11}\selectfont\ttfamily,
  keywordstyle=\bfseries\color{blue!40!black},
  commentstyle=\itshape\color{green},
  identifierstyle=\color{black},
  stringstyle=\color{orange},
  numbers=left,
  stepnumber=1,
  literate={\ \ }{{\ }}1
}

\lstdefinestyle{CustomJava}{
  belowcaptionskip=1\baselineskip,
  xleftmargin=12pt,
  xrightmargin=3pt,
  language=Java,
  numbersep=5pt,
  captionpos=b,
  tabsize=3,
  showstringspaces=false,
  basicstyle=\fontsize{9}{11}\selectfont\ttfamily,
  keywordstyle=\bfseries\color{purple!40!black},
  commentstyle=\itshape\color{blue},
  identifierstyle=\color{black},
  stringstyle=\color{cyan},
  numbers=left,
  stepnumber=1,
  literate={\ \ }{{\ }}1
}

\lstdefinestyle{CustomCS}{
  belowcaptionskip=1\baselineskip,
  xleftmargin=2pt,
  xrightmargin=3pt,
  language=C++,
  numbersep=5pt,
  captionpos=b,
  tabsize=3,
  showstringspaces=false,
  basicstyle=\fontsize{9}{11}\selectfont\ttfamily,
  keywordstyle=\bfseries\color{blue!40!green},
  commentstyle=\itshape\color{orange},
  identifierstyle=\color{red!40!blue},
  stringstyle=\color{red},
  numbers=left,
  stepnumber=1,
  literate={\ \ }{{\ }}1
}

\begin{document}
\maketitle

\begin{abstract}

Software developers write a lot of source code and documentation during software development. Intrinsically, developers often recall parts of source code or code summaries that they had written in the past while implementing software or documenting them. To mimic developers' code or summary generation behavior, we propose a retrieval augmented framework, \tool, that retrieves relevant code or summaries from a retrieval database and provides them as a supplement to code generation or summarization models. \tool has a couple of uniqueness. First, it extends the state-of-the-art dense retrieval technique to search for relevant code or summaries. Second, it can work with retrieval databases that include unimodal (only code or natural language description) or bimodal instances (code-description pairs). We conduct experiments and extensive analysis on two benchmark datasets of code generation and summarization in Java and Python, and the promising results endorse the effectiveness of our proposed retrieval augmented framework.

\end{abstract}

\section{Introduction}
\label{sec:into}
\begin{figure*}
\captionsetup[subfigure]{labelformat=empty}
\centering
\includegraphics[width=1.0\linewidth]{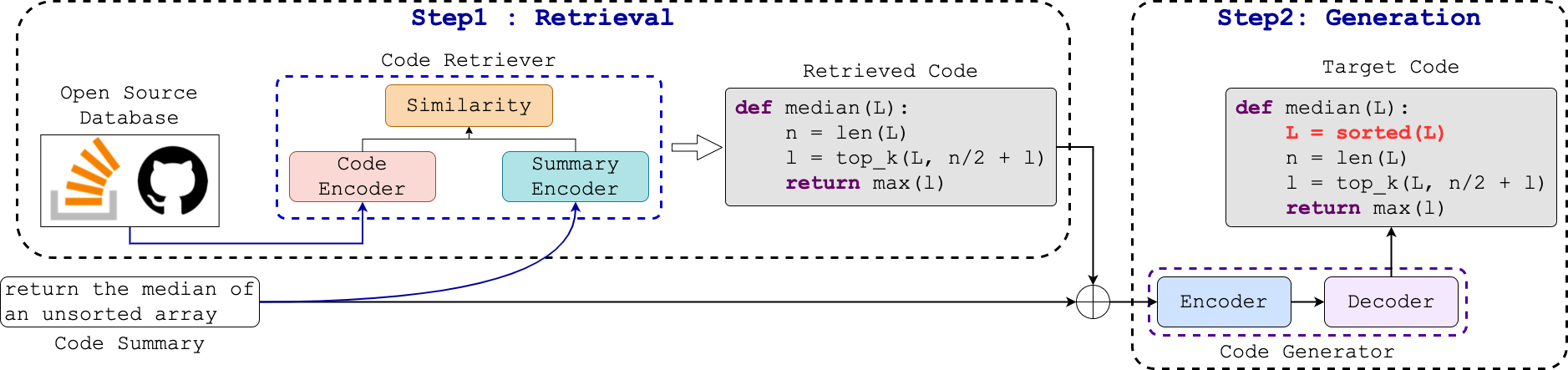}
\caption{Illustration of our proposed framework \tool for code generation. Given an input summary, we first retrieve top-$k$ candidate code ($k$=1 in this example).  We then aggregate them and based on that a \emph{generator} module  generates the target sequence.  }
\vspace{-2mm}
\label{figure:workflow}
\end{figure*}

In recent years, automating source code generation and summarization is receiving significant attention due to its potential in increasing programmers’ productivity and reducing developers’ tedious workload.
Consequently, various approaches have been explored in the literature to facilitate code generation \cite{yin-neubig-2017-syntactic, gu2016deep} and code documentation/summarization \cite{ahmad-etal-2020-transformer, wei2019code, allamanis2018survey}.
Despite initial success, most of the generated code still suffers from poor code quality~\cite{xu2021ide}. 
Therefore, the question remains---how to generate better code from a given summary and vice versa.

Source code generation and summarization, however, are intrinsically complex and challenging.
They involve generating diverse token sequences such as different variables, operators, keywords, classes, and method names~\cite{parvez-etal-2018-building}, which requires understanding the programming languages at lexical, syntax, and semantics levels. 
To combat these issues, recent studies (\eg \citet{ahmad2021unified, guo2020graphcodebert, xu-etal-2020-incorporating, feng2020codebert, xu-etal-2020-incorporating}) take a learning-based approach---they train representations of code and the associated text by leveraging existing high-quality source code and short text descriptions available in open-source repositories and question answering forums such as GitHub and Stack Overflow. 
Then fine-tune the representation models on the downstream tasks. 
Although these dataset contains high-quality human-written code and text, since the existing approaches do not directly leverage them during the generation process, the gain achieved by these approaches is still limited, especially when the source code is long.

To overcome this, we take advantage of the existing high-quality source code and their description by including them directly in the generation process that are retrieved via information retrieval technique.
In this work, we present \toolnospace, \textrm{a {\bf R}etrieval augment{\bf ED} {\bf COD}e g{\bf E}neration and summa{\bf R}ization} framework. 
While designing \tool, we take motivation from how developers take advantage of existing resources. For example, 
developers often search for relevant code 
in the code repository, and if found, adapt the retrieved code in their own context. 
Similarly, when an API usage is unclear, they search in question answering forums (\eg StackOverflow)~\cite{brandt2010example, sadowski2015developers}. Such an additional resource helps developers to increase their development productivity~\cite{li2013help}.

We design \tool as a two-step process (see \Cref{figure:workflow}). In the first step, given the input ({\em nl} text for code generation, or {\em code snippet} for summarization) a \emph{retriever} module retrieves relevant source code 
(for code generation) 
or summaries (for code summarization) from a database.\footnote{The database could be open source repositories (\eg GitHub) or developers' forums (\eg Stack Overflow).} In the second step, a \emph{generator} processes the retrieved code/summary along with the original input to generate the target output. In this way, \tool enhances the generation capability by augmenting the input through retrieval. 
The two-step process allows us to design a modular and configurable framework for source code and summary generation. Various designs of retriever and generator models can be incorporated into this framework.


Existing cross-encoder code retrievers being computationally expensive, their applicability to retrieve from a large database is limited~\cite{Humeau2020Poly-encoders:}. A natural choice would be to use sparse term based retrievers such as TF-IDF or BM25~\cite{robertson2009probabilistic}.  However, the \emph{retriever} module in \tool should exhibit a good understanding of source code and programmers' natural language, which is a non-trivial task due to the syntactic and semantic structure of the source code~\cite{guo2020graphcodebert, ahmad2021unified}. Such an expectation of searching for semantically similar code and summary may not be attainable by a sparse token level code retriever (\eg BM25). To that end, we design the {\em retriever} module in \tool based on programming languages (PL)  and natural languages (NL) understanding models (\eg GraphCodeBERT~\cite{guo2020graphcodebert}). This {\em retriever} module extends the state-of-the-art dense retrieval technique~\cite{karpukhin-etal-2020-dense} using two different encoders for encoding the query and document. 

As for the {\em generator}, \tool can handle retrieval databases consisting of both unimodal  (only code or natural language description) and bi-modal instances (code-description pairs) and makes the best usage of all the auxiliary information that are available. Yet, to incorporate information, we augment the retrieved  information only in the input level. It  does not modify the underlying architecture of the \emph{generator} module ---preserving its model agnostic characteristics.

\begin{figure}
\captionsetup[subfigure]{labelformat=empty}
\vspace{-5mm}
\centering
\includegraphics[width=1\linewidth]{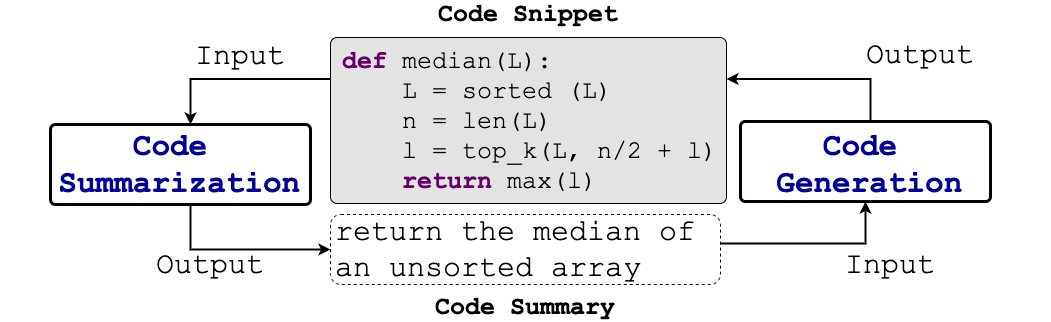}
\caption{Example input/output for the code generation and summarization tasks.}
\vspace{-3mm}
\label{figure:tasks}
\end{figure}

We evaluate the effectiveness of \tool on two popular programming languages (Java and Python) on both code generation and code summarization tasks. The empirical results show that, \tool's concept of {\em retrieval augmented generation} elevates the state-of-the-art code generation from an Exact Match score of 18.6 to 23.4 and the summary generation BLEU-4 score from 18.45 to 22.95 even when we forcefully remove the target candidate from the retrieved code or summary. With further experiments, we establish the importance of both the retrieved code and retrieves summary in the generation process.
The source code for reproducing our experiments are at \url{https://github.com/rizwan09/REDCODER}.

\section{Background}
\label{sec:background}
We first introduce the problem formulation and discuss the fundamentals of the {\em retriever} and {\em generator} components that \tool is built upon.

\subsection{Problem Formulation}
\label{sec:problem-formulation}
Our goal is two folds: (i) code generation: Generating source code ($C$), given their natural language description, such as code summaries, code comments or code intents ($S$); (ii) code summarization: Generating natural language summaries $S$, given source code snippets $C$. Fig \ref{figure:tasks} shows an example.

Let $X$ and $Y$ denote a collection of input and output sequences ($X = S_1, \ldots, S_n$, $Y = C_1, \ldots, C_n$ in code generation, $X = C_1, \ldots, C_n$, $Y = S_1, \ldots, S_n$ in summary generation
). 
We assume that we have access to a retrieval database consisting of an extensive collection of source code (\eg aggregated from GitHub or Stack Overflow) or summaries (\eg docstrings, code comments) ($Y_R$). Note that, target sequences ($Y$) may or may not be present in the retrieval database (${Y}_R$).
Now, given an input $x \in X$, a {\em retriever} retrieves the top-$k$ relevant output sequences from the database: $\mathcal{Y}_{1}, \mathcal{Y}_{2}, \ldots, \mathcal{Y}_{k} \in {Y}_R$.
Then the input sequence $x$ is augmented with the retrieved sequences to form $x' = x \oplus \mathcal{Y}_{1}  \oplus  \mathcal{Y}_{2}\ldots  \oplus \mathcal{Y}_{k}$, where $\oplus$ denote the concatenation operation.
Finally, a {\em generator} generates the target output $y \in Y$ given $x'$.
In the following, we first discuss the base {\em retriever} and {\em generator} modules used in \tool and then how we improve these components is in Section \ref{sec:method}. 


\subsection{Retriever: DPR}
\label{sec:background-retriever}
Information retrieval (IR) systems or retriever models
are designed to 
retrieve the top-$k$ relevant documents that presumably best provide the desired information~\cite{manning2008xml}. 
Term-based retrieval methods, \aka sparse retrieval models, such as TF-IDF or BM25 \cite{robertson2009probabilistic} 
use sparse vector representations to perform lexical matching and compute relevance scores to rank the documents based on a query.

On the other hand, dense retrieval methods encode documents into a fixed-size representations and retrieve documents via maximum inner product search \cite{NIPS2014_a14ac55a, guo2016quantization}. 
Particularly of interests, \citet{karpukhin-etal-2020-dense} propose a Dense Passage Retriever (DPR) model for open-domain question answering (QA). 
It consists of two encoders (\textbf{\it Q}(.) and \textbf{\it P}(.)) that encode queries and passages, respectively. 
The similarity of a query $q$ and a passage $p$ is defined by the inner product of their encoded vectors $sim(p,q)=Q(q)^T \cdot P(p)$.
Given a query $q$, a positive (relevant) passage $p^+$, and a set of $n$ irrelevant passages $p_i^-$, DPR optimizes the classification loss: 
\begin{equation*}
\setlength{\belowdisplayskip}{5pt}
\setlength{\abovedisplayskip}{5pt}
\label{eq:loss-ori-dpr}
    \begin{split}
    L = - \log \frac{e^{sim(q,p^+)}}{e^{sim(q,p^+)} + \sum_{i=1}^n e^{sim(q,p_i^-)}}.
 \end{split}
\end{equation*}

\citet{karpukhin-etal-2020-dense} propose to fine-tune DPR using \emph{in-batch negatives} ~\cite{gillick-etal-2019-learning, yih2011learning} with curated ``hard'' negatives using BM25 (candidates with high BM25 scores but contain no sub-string  that match the target). 
We refer to \citet{karpukhin-etal-2020-dense} for  details.


\subsection{Generator: PLBART }
\label{sec:background-plbart}



PLBART \cite{ahmad2021unified} is a sequence-to-sequence Transformer model \cite{vaswani2017attention} that is pre-trained on a huge collection of source code and natural language descriptions via denoising autoencoding.
PLBART has shown promise in several software engineering applications, including code generation and summarization. We adopt PLBART as the generator module in our proposed framework, \tool.


\begin{figure}[!t]
\captionsetup[subfigure]{labelformat=empty}
\centering
\includegraphics[width=1.0\linewidth]{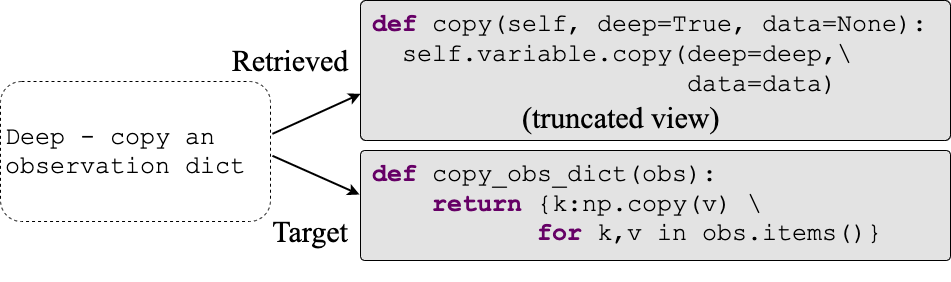}
\caption{An example  retrieved code that is relevant yet does not match the reference.}
\vspace{-2mm}
\label{figure:targetvsret}
\end{figure}

\section{Proposed Framework: \tool}
\label{sec:method}
Our proposed code generation and summarization framework, \tool
generates the target code or summary by augmenting the input $x$ with relevant code snippets or summaries. We build our {\em retriever} module by training a DPR model differently from \cite{karpukhin-etal-2020-dense}. With an intelligent scheme, we then augment the retrieved candidates and their pairs (if available) to  provide auxiliary supervision to the {\em generator}.  
We briefly describe the model components in this section.


\subsection{Retriever: \coder}
\label{sec:method:ret}
%
\paragraph{Architecture}
The {\em retriever} module of \tool is built upon the DPR model~\cite{karpukhin-etal-2020-dense} and we call it \coder (Summary and CODE Retriever). 
\coder composed of two encoders that encode source code and natural language summary.
We use bidirectional Transformer encoders \cite{vaswani2017attention} that are pre-trained on source code and natural language summaries.
Specifically, we explore CodeBERT \cite{feng-etal-2020-codebert} and GraphCodeBERT \cite{guo2020graphcodebert} as the code and summary encoders for \coder.



\paragraph{Input/Output} 

\coder takes an input sequence $x$ (code or summary) and retrieves a set of relevant documents from a database of output sequences $Y$ (if the input is code, then the output is summary and vice versa).
\coder returns the the top-$k$ output sequences $\{\mathcal{Y}_{1}, \mathcal{Y}_{2}, \ldots , \mathcal{Y}_{k}\}$, where $sim(x, \mathcal{Y}_{i}) \geq sim(x, \mathcal{Y}_{j}) \forall j>i$.

\begin{figure}[!t]
\captionsetup[subfigure]{labelformat=empty}
\centering
\includegraphics[width=1.0\linewidth]{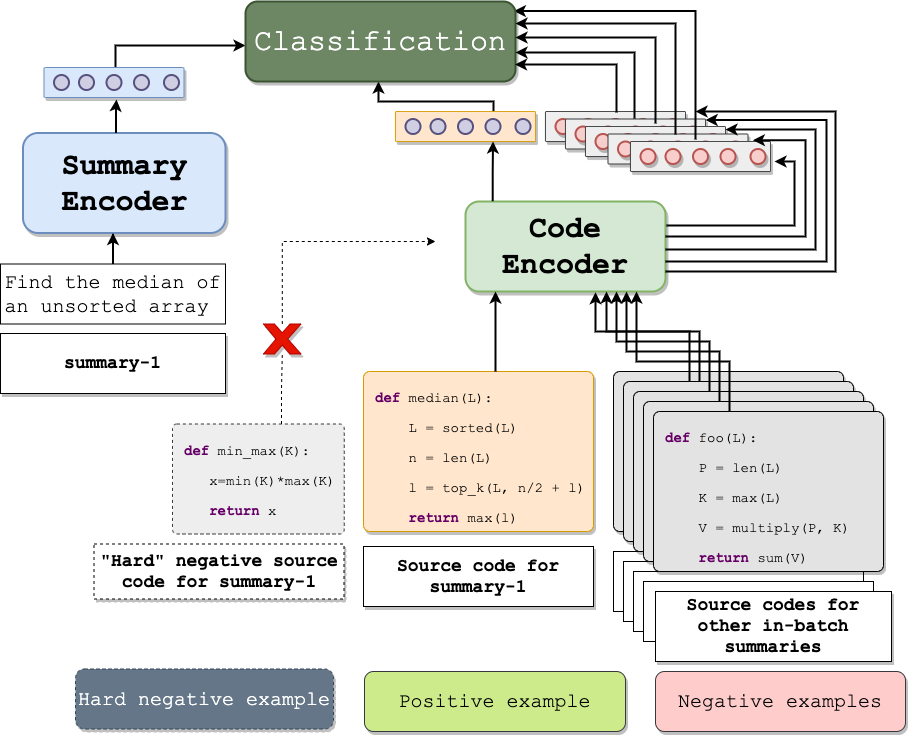}
\caption{
Training scheme of the {\em retriever} module (\coder) of our proposed framework REDCODER  for the code generation task. Unlike in open-domain QA \cite{karpukhin-etal-2020-dense}, we do not use ``hard'' negatives (\eg candidates retrieved by BM25 that do not exactly match the reference) during fine-tuning.
}
\vspace{-2mm}
\label{figure:dpr_train}
\end{figure}

\paragraph{Training}

We fine-tune \coder using a set of parallel examples ($x_i, y_i$) of code and summaries.
As mentioned in Section \ref{sec:background-retriever}, DPR originally proposed to be fine-tuned using \emph{in-batch negatives} and curated ``hard'' negatives from BM25 retrieved passages for open-domain QA. 
The key idea behind ``hard'' negatives is to fine-tune DPR to distinguish the target passage from relevant passages that do not contain the target answer.
However, unlike open-domain QA, a retrieved code or summary that is not the target could still benefit code generation or summarization (verified in Section \ref{sec:analysis}).
We provide an example in Figure \ref{figure:targetvsret}; although the retrieved code does not match the target one but can facilitate generating it.
Therefore, we fine-tune \coder without any ``hard'' negatives. 
Specifically, for each training instance ($x_i, y_i$), the corresponding output $y_i$ is considered as positive and the other in-batch outputs (\ie the outputs of other instances in the same batch - $y_1, \ldots, y_{i-1}, y_{i+1}, \ldots, y_{bsz}$) as negatives.
Figure \ref{figure:dpr_train} shows an example of \coder fine-tuning for code generation task. 

\begin{figure}[t]
\captionsetup[subfigure]{labelformat=empty}
\centering
\hspace*{-3pt}
\includegraphics[width=1.0\linewidth]{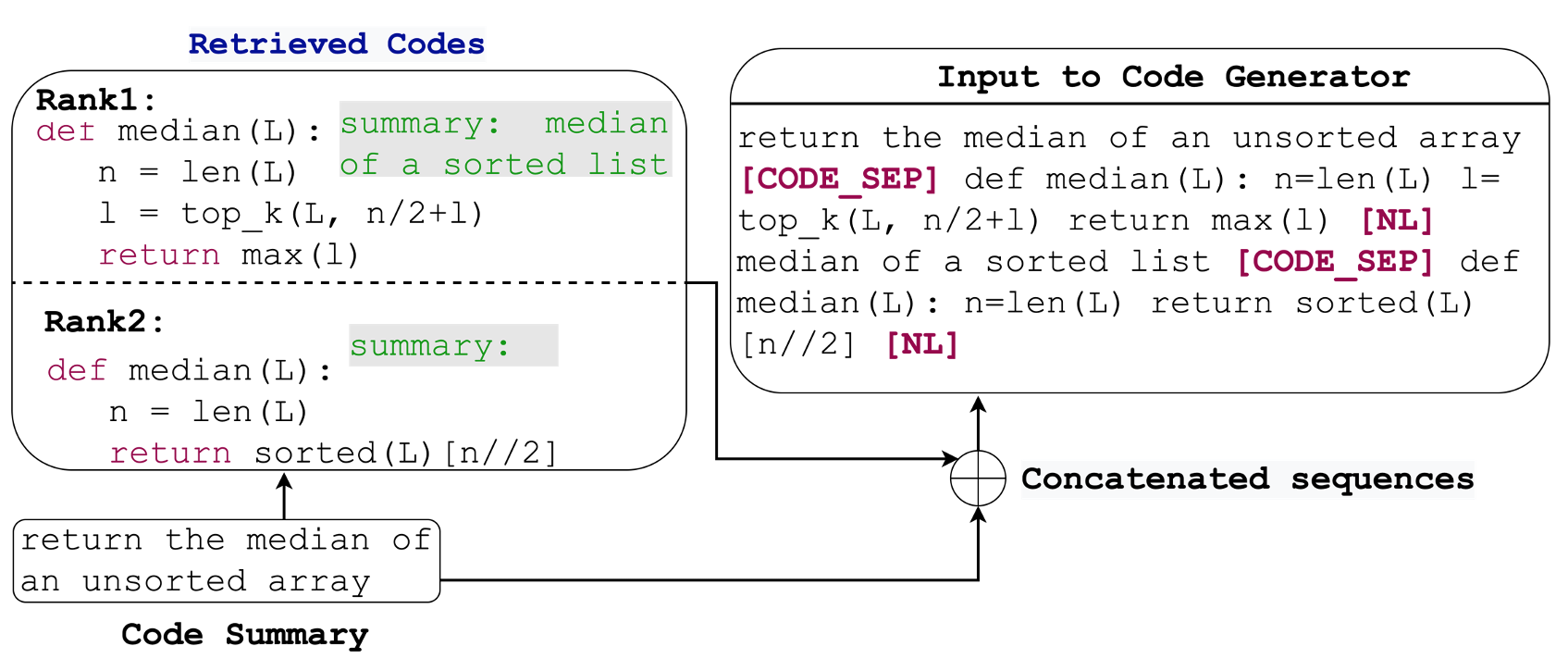}
\caption{\toolext input for code generation. }
\vspace{-2mm}
\label{figure:gen_input}
\end{figure}

\subsection{Generator: SCODE-G}
\label{sec:method:gen}
We adopt PLBART as discussed in Section \ref{sec:background-plbart} as the {\em generator} module of \tool and call it SCODE-G (Summary and CODE Generator).
The input sequence $x$ is concatenated with the top-$k$ retrieved sequences 
to form the augmented input sequence, $x' = x \oplus \mathcal{Y}_{1}  \oplus  \mathcal{Y}_{2} \ldots \oplus \mathcal{Y}_{k}$.
The augmented input $x'$ is fed to PLBART to estimate $p_{gen}(y | x')$.

Note that a source code often consists of docstrings, comments that can be extracted to form code -- summary pairs.
In the retrieval databases, code and summaries are either singleton (\eg code without a description or a problem statement without any code) or parallel.
Therefore, we consider two retrieval settings that require separate modeling consideration for the generator.




\begin{table*}[!ht]
\centering
    \resizebox{\linewidth}{!}{
        \begin{tabular}{l|c|c|c|c|c|c|c|c}
            \hline
            Dataset & Gen. & Sum. & Lang. & Train & Valid & Test & $|$Code$|$ & $|$Summary$|$ \\
            \hline
           CodeXGLUE  & \multirow{2}{*}{\cmark}  & \multirow{2}{*}{\cmark} & Java & 164,923 & 5,183 & 10,955 & 97 & 12	 \\
            \cite{CodeXGLUE}&  & & Python & 251,820 & 13,914 &  14,918 & 99 & 14	 \\
            \hline
            Concode~\cite{iyer-etal-2018-mapping} & \cmark & \xmark &  Java & 100,000 & 2,000 & 2,000 & 27 & 72 \\
            \hline
        \end{tabular}
        }
        \caption{Dataset Statistics. Gen., and Sum. refers to code generation and summarization tasks respectively. 
        Summary denotes a natural language description paired with each code.
        For Concode, the input summary includes the corresponding environment variables and methods. All lengths are computed and averaged before tokenization.}
        \label{tab:dataset_task}
    \vspace{-2mm}
\end{table*}

\paragraph{Case 1: Retrieve candidates are singleton} In this case, we concatenate the original input sequence $x$ and the top-$k$ retrieved candidates with a special separator token.
\begin{equation*}
\setlength{\belowdisplayskip}{5pt}
\setlength{\abovedisplayskip}{5pt}
\label{eq:common}
 \begin{split}
    x' =& \ x\  [csep] \ \mathcal{Y}_{1} \ [csep] \ \mathcal{Y}_{2} \ \ldots \ [csep] \ \mathcal{Y}_{k}.
 \end{split}
\end{equation*}
This is our default setting and we refer this as \tool in this work.


\paragraph{Case 2: Retrieve candidates are pairs} In this case, retrieved candidates are pair of code and natural language (NL) summary. We augment the input sequence using both of them as follows.
\begin{equation*}
\setlength{\belowdisplayskip}{5pt}
\setlength{\abovedisplayskip}{5pt}
\label{eq:common}
    \begin{split}
        x' =& 
        \ x \ [csep] \ \mathcal{Y}_{1} \ [nsep] \ \mathcal{X}_{1} \ [csep] \ \mathcal{Y}_{2} \\
        & \ [nsep] \ \mathcal{X}_{2} \ \ldots \ [csep] \ \mathcal{Y}_{k} \ [nsep] \ \mathcal{X}_{k},
    \end{split}
\end{equation*}
where $\mathcal{X}_{j}$ and $\mathcal{Y}_{j}$ are parallel sequences (\eg $\mathcal{Y}_{j}$ is a piece of code and $\mathcal{X}_{j}$ is its corresponding summary for the code generation task) retrieved from the database.
We conjecture that the additional information $\mathcal{X}_{j}$ complements the input sequence $x$ and verify its effectiveness in the experiments.

Note that retrieve candidates could be a mix of singleton and pairs. In case of a singleton candidate, we simply replace $\mathcal{X}_{j}$
or $\mathcal{Y}_{j}$ 
with an empty string.
We refer this setting as \toolextnospace. 
Although, \toolext is a more general setting which includes ``Case 1'', we study them separately to understand how these two retrieval settings benefit the target tasks.
We illustrate an example on code generation in Figure \ref{figure:gen_input}.
In both cases, the augmented input $x'$
is truncated to match PLBART's maximum input length 512.


\section{Experiment Setup}
\label{sec:exp_setup}


\begin{table*}[h]
\centering
\begin{tabular}{l|l|c c c|c c c}
\hline
\multicolumn{2}{c|}{Method} & \multicolumn{3}{c|}{Java}& \multicolumn{3}{c}{Python}\\
\hline
Type & Name & EM & BLEU & CodeBLEU & EM & BLEU & CodeBLEU \\
\hline
Retrieval & BM25 &  0.00 & 4.90 & 16.00 & 0.00 & 6.63 & 13.49 \\
Based & \coder  & 0.00 & 25.34  & 26.68 & 0.00 & {22.75} &  23.92  \\
\hline
\multirow{4}{*}{Generative} & CodeBERT & 0.00 & 8.38 & 14.52 & 0.00 & 4.06 & 10.42\\
& GraphCodeBERT & 0.00 & 7.86 & 14.53 & 0.00 & 3.97 & 10.55\\
& CodeGPT-adapted & 0.00 & 7.10 & 14.90 & 0.01 & 3.11 & 11.31 \\ 
& PLBART &  0.00 & 10.10 & 14.96 & 0.00 & 4.89  & 12.01  \\
\hline
Retrieval & BM25 + PLBART  & 0.10 & 11.37 & 15.52 & 0.03 & 6.99   & 13.89 \\
Augmented & \tool   & {8.95} & { 26.92}  & {31.15} & { 8.88} & { 22.74}  & {28.93}   \\
Generative & \toolext & {\bf 10.21} & {\bf 28.98}   & {\bf 33.18} & {\bf 9.61} & {\bf 24.43} &  {\bf 30.21} \\
\hline 
\end{tabular}
\caption{
Results on code generation on CodeXGLUE \cite{CodeXGLUE}.
}
\label{table:csnet_gen}
\vspace{-2mm}
\end{table*}

In order to investigate the effectiveness of our framework, we perform a comprehensive study and analysis on code generation and summarization in two programming languages, Java and Python.

\subsection{Datasets and Implementations}

\noindent\textbf{Datasets\hspace{0.5em}} 
We perform evaluation on both the tasks using the code summarization dataset from CodeXGLUE~\cite{CodeXGLUE}.
It is curated from  CodeSearchNet \cite{husain2019codesearchnet} by filtering noisy examples.
In addition, we conduct code generation experiments in Java using the Concode benchmark~\cite{iyer-etal-2018-mapping}.
The dataset statistics are summarized in Table \ref{tab:dataset_task}.

\smallskip\noindent\textbf{Retrieval Databases\hspace{0.5em}} 
To generate a source code given its natural language description or a summary given the code, our proposed approach \tool first retrieves prospective candidates from an existing code or summary database. 
We form the code retrieval database using the deduplicated source code (on average 1.4M functions in Java and Python) that consists of both paired (59\%) and monolingual code, released in CodeSearchNET~\cite{husain2019codesearchnet}. As for building the summary retrieval database, we extract the high quality natural language summaries from the paired instances in the training sets  of CodeSearchNET. As many of the summaries are duplicated, we also consider the training sets in the other four available languages Ruby, Javascript, Go, and PHP. We then further enlarge it by aggregating the additional summaries from the CCSD corpus \cite{liu2021retrievalaugmented}.
After performing deduplication, we retain 1.1M unique code summaries and for evaluating \toolext, 20\% of them can be used as pairs with the corresponding Java and Python source code. We provide the statistics of the retrieval databases in Appendix.
Note that the retrieval databases contain code and summaries that are curated from real developers' open sourced repositories on GitHub. 
By default, we exclude the target code/summary from the retrieval database. 

\smallskip\noindent\textbf{Implementations\hspace{0.5em}} 
As mentioned in Section \ref{sec:method}, \tool has two disjoint components.
First, the dense retriever \coder is implemented adopting DPR~\cite{karpukhin-etal-2020-dense} and the encoders in DPR are initialized from  GrpahCodeBERT available in the Huggingface API~\cite{wolf-etal-2020-transformers}. 
In addition, we implement a baseline BM25 retriever.
We use the official codebase of PLBART~\cite{ahmad2021unified} and set max epoch to 15, patience to 5, learning rate to $2\times 10^{-5}$.
We tune the batch size in \{8, 16, 32, 64, 72\} and the $k$ value for top-$k$ retrieval up to 10  for code generation and in range \{10, 30, 50, 100\} for code summarization. 
As some candidate code and summaries are short in length, we tune with this upper bound of $k$ to accommodate as many candidates as possible within PLBART's maximum input length.


\subsection{Evaluation Metrics}
\noindent\textbf{BLEU\hspace{0.5em}} 
Following prior works \cite{ahmad2021unified, feng2020codebert}, we compute the corpus level BLEU \cite{papineni-etal-2002-bleu} and the smoothed BLEU-4 \cite{lin-och-2004-orange} scores for code generation and summarization tasks.


\smallskip
\noindent\textbf{{CodeBLEU}\hspace{0.5em}} 
To demonstrate syntactic and semantic data flow correctness of code generation models, we report CodeBLEU \cite{ren2020codebleu}.
CodeBLEU is a weighted average of lexical, abstract syntax tree, and data flow match.



\smallskip
\noindent\textbf{{Exact Match (EM)}\hspace{0.5em}} 
indicates the percentage of output sequences that exactly match the references.


\begin{table}[t]
\centering
\resizebox{\linewidth}{!}{%
\begin{tabular}{l|c c c}
\hline
Methods & EM & BLEU & CodeBLEU \\ 
\hline 
\multicolumn{4}{l}{Retrieval based methods}\\
\hline
BM25 &  0.0 & 20.3 & 23.7  \\
\coder &  0.0 &  32.6 & 36.5\\
\hline 
\multicolumn{4}{l}{Generative methods}\\
\hline
Seq2Seq & 3.1 & 21.3 & 26.4 \\
\citet{guo-etal-2019-coupling} & 10.1 & 24.4 & 29.5 \\
\citet{iyer-etal-2019-learning} & 12.2 & 26.6 & - \\
GPT-2 & 17.4 & 25.4 & 29.7 \\
CodeGPT-2 & 18.3 & 28.7 & 32.7 \\
CodeGPT-adapted & { 20.1} & 32.8 & 36.0 \\ 
CodeBERT & 18.0 & 28.7 & 31.4 \\
GraphCodeBERT & 18.7 & 33.4 & 35.9 \\
PLBART & 18.6 & {36.7} & {38.5} \\
\hline 
\multicolumn{4}{l}{Retrieval augmented generative methods}\\
\hline
BM25+PLBART &  21.4 & 40.2 & 41.8\\
\tool & {\bf 23.4} & { 41.6}  & {\bf 43.4} \\
\toolext & {23.3} & {\bf 42.5}  & {\bf 43.4} \\
\hline
\end{tabular}
}
\caption{ 
Code generation results on Concode dataset. 
\coder was initialized with CodeBERT. GraphCodeBERT initialized results are similar. 
}
\label{table:concode}
\vspace{-2mm}
\end{table}
\begin{table}[t]
\centering
\begin{tabular}{l|c | c}
\hline
Methods & Python & Java  \\
\hline
\multicolumn{3}{l}{Retrieval based methods} \\ \hline
BM25 & 1.92 & 1.82 \\
\coder  & 14.98 & 15.87 \\ \hline
\multicolumn{3}{l}{Generative methods} \\ \hline
Seq2Seq &   15.93	& 15.09	 \\
Transformer & 15.81	& 16.26	\\
RoBERTa & 18.14 & 16.47  \\
CodeBERT & 19.06 & 17.65  \\ 
GraphCodeBERT & 17.98 & 17.85 \\
PLBART &  { 19.30} & {18.45}  \\ \hline
\multicolumn{3}{l}{Retrieval augmented generative methods} \\
\hline
BM25 + PLBART & 19.57 & 19.71 \\
\tool & {\bf 21.01} & { 22.94} \\
\toolcode & { 20.91} & {\bf 22.95} \\
\hline
\end{tabular}
\caption{Evaluation BLEU-4 score for code summarization on CodeXGLUE. Baseline results are reported from \citet{ahmad2021unified}.}
\label{table:code_to_text}
\vspace{-2mm}
\end{table}


\begin{table*}[!ht]
\centering
\resizebox{\linewidth}{!}{%
\begin{tabular}{ l | c@{\hskip 0.1in} c@{\hskip 0.1in} c| c@{\hskip 0.1in} c@{\hskip 0.1in} c| c@{\hskip 0.1in} c@{\hskip 0.1in} c }
\hline
\multirow{ 2}{*}{Methods} & \multicolumn{3}{c|}{CodeXGLUE (Java)} &  \multicolumn{3}{c|}{CodeXGLUE (Python)} & \multicolumn{3}{c}{Concode (Java)}  \\ 
\cline{2-10}
& BLEU & EM & CodeBLEU & BLEU & EM & CodeBLEU  & BLEU & EM & CodeBLEU\\ 
\hline
 \coder & {36.6} & 21.0 & 37.9 & {35.6} & {19.2} & 35.1 &  70.3 & 61.7 & 72.0 \\
\tool  & 36.3 & {29.4} & { 41.4} & { 32.1} & { 27.5} & {38.0 } &  { 76.7} & { 67.5} & { 76.5}  \\
\toolext  & {\bf 42.8} & {\bf 37.0}  & {\bf 47.3} &  {\bf 38.9} & {\bf 34.5} & {\bf 43.8} &  {\bf 81.7} & {\bf 76.2} & {\bf 81.7} \\
\hline

\end{tabular}
}
\caption{
Results on code generation keeping the target code in the retrieval database.  
}
\label{tab:gen_improve_with_ref}
\vspace{-2mm}
\end{table*}

\begin{table}[t]
\centering
\begin{tabular}{l|l|c|c}
\hline
Settings & Methods & Python & Java \\
\hline
 & RoBERTa & 0.587	& 0.599 \\
Cross- & RoBERTa (code) & 0.610 & 0.620 \\
Encoder & CodeBERT & 	0.672 & 	0.676 \\
& GraphCodeBERT & {\bf 0.692} & {\bf 0.691} \\
\hline
\multirow{2}{*}{Bi-} & DPR & 0.093 & 0.064 \\
& DPR (code) & 0.398 & 0.462 \\
\multirow{-2}{*}{Encoder}& \coder & 0.690 & 0.686\\
\hline
\end{tabular}
\caption{
MRR results on code retrieval from the validation and test set in CodeXGLUE. Our bi-encoder retriever \coder is comparable with other cross-encoder models while it is much faster. DPR refers to \citet{karpukhin-etal-2020-dense} and DPR (code) is trained with BM25 ``hard'' negative training schema built upon our source code datasets.
}
\label{table:code_mrr}
\vspace{-2mm}
\end{table}

\subsection{Baseline Methods}
We compare \tool \wrt a number of  state-of-the-art code models. We classify them into two categories: (i) retrieval based models and (ii) generative models. 
We study both generative models that are trained from scratch and are pre-trained on programming and natural languages.


\paragraph{Retrieval based models}
We examine two retriever baselines and consider the top-1 retrieved candidate as the prediction.

\noindent$\bullet$~\textbf{Dense Retriever\hspace{0.5em}} 
We consider DPR as the dense retriever baseline. 
We evaluate both the officially released models trained on the natural language open-domain QA task and a variant called DPR (code) that we fine-tune on the evaluation datasets.


\noindent$\bullet$~\textbf{Sparse Retriever\hspace{0.5em}}
The second baseline is a sparse retriever that uses the BM25 algorithm to compute relevance scores.


\vspace{-1mm}
\paragraph{Generative models}
The generative models work in a sequence-to-sequence (Seq2Seq) fashion.

\noindent$\bullet$~\textbf{RoBERTa, RoBERTa (code)\hspace{0.5em}} 
RoBERTa  models~\cite{liu2019roberta} pre-trained on natural language corpora, and source code from CodeSearchNet~\cite{husain2019codesearchnet} respectively.

\noindent$\bullet$~{\textbf{CodeBERT}~\cite{feng2020codebert} \hspace{0.5em}} 
is pretrained with a hybrid objective incorporating masked language modeling~\cite{devlin2018bert} and replaced token detection~\cite{clark2020electra}. 

\noindent$\bullet$~{\textbf{GraphCodeBERT}~\cite{guo2020graphcodebert} \hspace{0.5em}} is pre-trained by modeling the data flow graph of source code. GraphCodeBERT holds the state-of-the-art results on code search using CodeSearchNet.


\noindent$\bullet$~\textbf{GPT-2, CodeGPT-2, and CodeGPT-adapted \hspace{0.5em}}
are GPT-style models that are pre-trained on natural language~\cite{radford2019language} and code corpora CodeXGLUE~\cite{CodeXGLUE}.

\noindent$\bullet$~{\textbf{PLBART}~\cite{ahmad2021unified}\hspace{0.5em}} is the generator module of our proposed framework.

In addition, we train an LSTM based Seq2Seq model with attention mechanism~\cite{luong-etal-2015-effective} and a Transformer model \cite{vaswani2017attention} on the benchmark datasets. 




\section{Results}

    \subsection{Code Generation}
    \label{sec:exp-result-codegen}

    \Cref{table:csnet_gen} and \Cref{table:concode} show the evaluation results on code generation from summary descriptions on CodeXGLUE, and Concode datasets, respectively. First, we compare \tool with the state-of-the-art code generation models. They are transformers models pre-trained with different objectives using external resources of different sizes. Among them, the relatively strong baseline PLBART has an EM score of 18 on the Concode dataset while it rarely generates any code that matches the real target code in CodeXGLUE (See Table \ref{table:csnet_gen}) (more discussion on this is in Appendix).  The BLEU and CodeBLEU scores are also low. Such result indicates that automated code lacks quality and correctness without the proper supervision in the input to the generator. 
    
    Among the retriever-only models, \coder significantly outperforms BM25 (more comparison is in \cref{sec:analysis}). As expected, the EM is zero as targets are filtered from the retrieval, and CodeBLEU scores are high as they are real code. However, although the retrieved code does not exactly match the target code, they are quite relevant (\eg Figure \ref{figure:targetvsret}; more in Appendix). When comparing retrieval-only models to generative models, it is interesting to note that \coder surpasses PLBART by a large margin on CodeXGLUE (\Cref{table:csnet_gen}), suggesting that retrieved code has high overlapping with target code that can benefit the generation.
    
    Overall, the retrieval augmented generative models excel in code generation. Our proposed framework \tool outperforms PLBART by a large margin, validating the advantage of reusing existing codebases to help code generation.  The \toolext gains are even higher. For CodeXGLUE (Java, Python) and Concode, the  gains in BLEU are 18.88, 19.54, and 5.8. Comparing \tool to \toolext shows that BLEU scores on Concode and all metrics on CodeXGLUE are improved by $\sim$1\%. These results confirm our conjecture that complementing input with paired summaries of the retrieved code help code generation.  We provide a qualitative example in the Appendix to explain how the retrieved information helps PLBART in generation.

    \subsection{Code Summarization}
    \label{sec:exp-result-codesum}
    
    We compare \tool with three sets of baseline methods for code summarization, and \Cref{table:code_to_text} shows the results.  
    Among the two retrieval base methods, \coder performs significantly well, confirming the advantages of dense retrieval over its sparse counterpart.
    Out of the generative methods, PLBART excels on code summarization as it leverages an extensive collection of natural language descriptions during pre-training.
    As anticipated, retrieval augmented generative methods outperform the other two sets of models.
    We see that the ``BM25 + PLBART'' model improves over PLBART, confirming our conjecture that retrieval augmented techniques have the promise to improve code summarization.
    Our proposed framework \tool and its variant \toolext outshine ``BM25 + PLBART'', surpassing its performance by $\sim$1.5 and $\sim$3.2 points for Python and Java languages, respectively.
    
\section{Analysis}
\label{sec:analysis}
In this Section, we analyze REDCODER's performance on the following points.

\begin{figure}[t]
\centering
\includegraphics[width=0.6\linewidth]{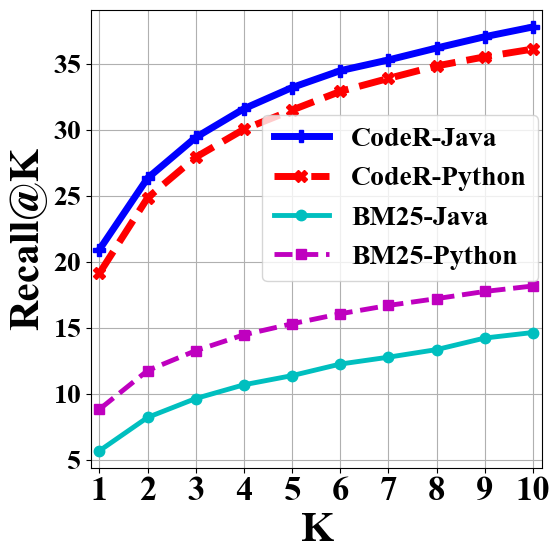}
\caption{ Recall@K for  CodeR and BM25. CodeR refers to \coder used for source code retrieval.} 
\vspace{-2mm}
\label{fig:recall-bm25-coder}
\end{figure}

\smallskip\noindent\textbf{Retrieval database includes the target sequence\hspace{0.5em}}
\Cref{tab:gen_improve_with_ref} shows the code generation  results when we did not filter the target from the retrieval (summarization results are in Appendix). As expected, \coder performances are much  better than those in Table \ref{table:csnet_gen}, \ref{table:concode}, and \ref{table:code_to_text}. In all cases, \tool gets more enhanced when target is present in the retrieval database. 
For the code generation task, we plot the recall@k curve for $k$ upto 10 for both Java and Python on CodeXGLUE dataset when the retrieval contains the target in Figure \ref{fig:recall-bm25-coder}. As we can see,  \coder significantly outperforms in both languages and for all $k$ values.

\smallskip\noindent\textbf{Bi-encoder \coder vs cross-encoder retrievers\hspace{0.5em}}
\Cref{table:code_mrr} shows the retrieval performance of different alternative retrieval techniques that we considered in \tool. \coder performs comparably well with GraphCodeBERT while being significantly faster and scalable~\cite{Humeau2020Poly-encoders:}. Note that, \coder also uses GraphCodeBERT to initialize its encoders (see \Cref{figure:dpr_train}). However, \coder's design of using different encoders for query and documents enables pre-indexing of database and faster retrieval in practice.  

\begin{figure}[t]
\centering
\vspace{-3mm}
\includegraphics[width=1.0\linewidth]{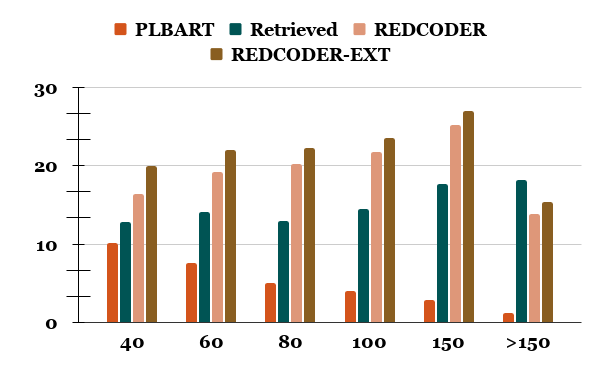}
\caption{(Python) Code gen. BLEU vs target len. }\label{fig:bleu_vs_len_python}
\end{figure}

\smallskip\noindent\textbf{Performance vs target length\hspace{0.5em}}
\Cref{fig:bleu_vs_len_python} shows the code generation performances of different models \wrt the target code length for Python. While the generator model (PLBART)'s performance consistently decreases with increasing code size, the retriever (\coder) performs consistently well. Such consistent performance from \coder boosts performance of \tool (and also \toolext) {\em significantly higher} than the generative model counterpart. For Java, we find similar results (details in Appendix).



\begin{table*}[!htb]
\centering
\begin{tabular}{l| c c c |  c c c }
\hline
\cline{2-7}
\multirow{2}{*}{Model} &  \multicolumn{3}{c|}{Human Evaluation} & \multicolumn{3}{c}{Automatic Metric} \\
\cline{2-7}
& Similarity & Relevance & Compilability & BLEU & EM & CodeBLEU \\
\hline
\coder & 2.09 & 3.00 & 3.16 & 11.56 & 0.00 & 16.66 \\
  \tool & 2.06 & 2.94 & 3.10 & 10.70 & 0.07 & 18.31 \\

\hline

\end{tabular}
\caption{
Human evaluation on code generation (CodeXGLUE-Python). \tool (\coder + SCODE-G)  achieves similar scores as \coder that directly retrieves developers' written code which suggests that the quality of the code generated by SCODE-G are competitive with real code from programmers’ perspective.
}
\label{table:hum_eval}
\vspace{-2mm}
\end{table*}

\begin{figure}[t]
    \centering
    \begin{subfigure}[b]{0.49\linewidth}
        \centering
        \includegraphics[width=\linewidth]{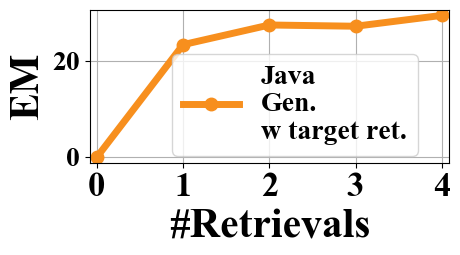}
        \caption[]%
        {{\small CodeXGLUE (Java) gen.}}    
        \label{fig:em_vs_num_retrivals_java_gen}
    \end{subfigure}
    \hfill
    \begin{subfigure}[b]{0.49\linewidth}  
        \centering 
        \includegraphics[width=\linewidth]{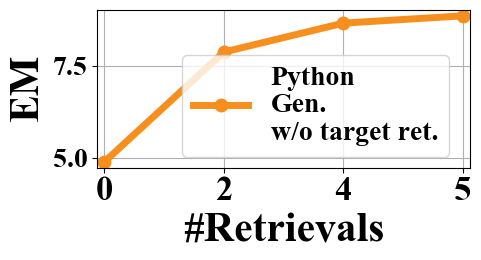}
        \caption[]%
        {{\small CodeXGLUE (Python) gen.}}    
        \label{fig:em_vs_num_retrivals_python_gen}
    \end{subfigure}
    \vskip\baselineskip
    \begin{subfigure}[b]{0.49\linewidth}   
        \centering 
        \includegraphics[width=\linewidth]{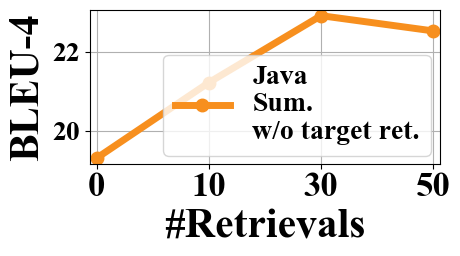}
        \caption[]%
        {{\small CodeXGLUE (Java) sum.}}    
        \label{fig:em_vs_num_retrivals_java_sum}
    \end{subfigure}
    \hfill
    \begin{subfigure}[b]{0.49\linewidth}   
        \centering 
        \includegraphics[width=\linewidth]{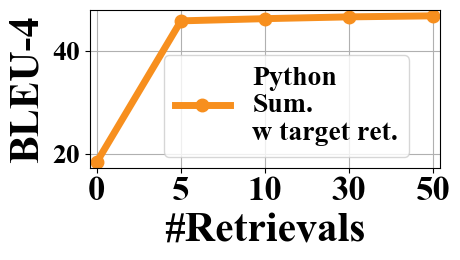}
        \caption[]%
        {{\small CodeXGLUE (Python) sum.}}    
        \label{fig:em_vs_num_retrivals_python_sum}
    \end{subfigure}
    \caption{Code gen. and sum. performance vs \#retrievals. In general performance improves with higher number of augmented candidates. } 
    \vspace{-2mm}
    \label{fig:per_vs_num_retrivals}
\end{figure}

\smallskip\noindent\textbf{Performance vs \#retrievals\hspace{0.5em}}
Figure \ref{fig:per_vs_num_retrivals} shows that typically the performance improves more with more retrievals on both tasks. However, roughly 5 code and 30 summaries work sufficiently well.


\smallskip\noindent\textbf{Human evaluation \hspace{0.5em}}
Finally, we evaluate the quality of code generated by SCODE-G using human evaluation. 
In Table \ref{table:hum_eval}, we perform a human evaluation for code generation task on a subset of the test set in CodeXGLUE (Python). In this study, we compare \tool generated code with the code retrieved by \coder. Note that both \tool and \coder using the same retrievers, but \tool generates code using SCODE-G, while \coder outputs code written by real programmers. 
We sample 30 instances where \tool generated code has a lower BLEU score than that of the \coder and investigate whether the quality of code generated by them are significantly different on these cases. 

As programming requires a specific skill, we do not evaluate the quality of the code generation using the mass crowd workers.
We recruit 7 Ph.D. students studying in computer science as volunteers\footnote{Before participating in the evaluation process, all the participants are informed that it is a voluntary task and it may take roughly 30 minutes to perform the evaluation.} to score (1 to 5) code based on three criteria (i) similarity, and (ii) relevance \wrt the target code; (iii) the compilability of the generated code.

The ratings show that both models receive similar scores, with a slightly higher score for \coder in terms of similarity to the target code, relevancy, and compilability. This shows that the quality of the code generated by SCODE-G are competitive with real code from programmers' perspective. Interestingly, \tool achieves higher scores than \coder in 
 CodeBLEU and Exact Match even on the cases where its BLEU score is lower. 

\section{Related Works}

\noindent\textbf{Code Summarization.} In recent years, source code summarization attracted a lot of attention 
~\cite{iyer2016summarizing, liang2018automatic,allamanis2016convolutional,hu2018summarizing,ahmad-etal-2020-transformer}. Many of these works view code as a sequence of token.
Other approaches leverage the structural properties of code using Tree based model~\cite{shido2019automatic, harer2019tree, hu2018deep, leclair2019subroutines}.
In literature, several retrieval-based methods were proposed that leverage retrieved information along with the input code. 
For example, \citet{zhang2020retrieval} retrieves similar code snippet and use those as an auxiliary input for summarization. 
On the other hand,  \citet{hayati-etal-2018-retrieval} retrieves related summaries for augmenting summarization input. Different from these approaches, \tool leverages both the retrieved code and its summary to augment the input. 

\smallskip
\noindent\textbf{Code Generation.} Generating source code is a major stepping stone towards automated programming.
\citet{yin-neubig-2017-syntactic}, and \citet{rabinovich-etal-2017-abstract} proposed code generation as abstract syntax tree generation to ensure its syntactic correctness.
Recent advancements in pre-training language models on unlabeled source code data \cite{CodeXGLUE, ahmad2021unified} showed colossal promise towards learning code syntax and semantics, resulting in improved code generation models.

\smallskip
\noindent\textbf{Code Retrieval and Others.} 
Numerous software engineering applications require information retrieval. \citet{sadowski2015developers, xia2017developers, stolee2014solving, sim2011well} show that developers search for related code, API examples for implementing or adapting new APIs.  Design of \tool is inspired by developers' behavior while writing code. Developers use search engines for retrieving off-the-shelf libraries~\cite{hucka2018software}, or ``usable'' source code~\cite{rahman2018evaluating} for adapting in the development process~\cite{nasehi2012makes, arwan2015source, ponzanelli2014mining}. Similarly, \tool retrieves existing code or summaries and adapts them to generate the target code or summary. In contrast, \citet{NEURIPS2018_cd17d3ce} optimizes a joint objective; \citet{zhang2020retrieval,liu2021retrievalaugmented} do not consider  any decoder pre-training, \citet{RAG_NEURIPS2020_6b493230} fine-tunes both of the retriever and the generator end-to-end. For open domain QA, \citet{izacard-grave-2021-leveraging} propose a similar model of alternative generator (multi-encoder uni-decoder). 


\section{Conclusion}



We propose \tool to automate developers' writing of code and documentation by reusing what they have written  previously. 
We evaluate \tool on two benchmark datasets and the results demonstrate a significant performance boost with the help of the retrieved information. In the future, we want to extend \tool to support other code automation tasks such as code translation.

\section*{Acknowledgments}
We thank anonymous reviewers for their helpful feedback.
We also thank the UCLA NLP group for helpful discussions, comments, and participating voluntarily in the human evaluation.
This work was supported in part by NSF 
OAC-1920462, 
SHF-2107405, SHF-1845893, IIS-2040961, IBM, and VMWare.
Any opinions, findings, and conclusions expressed herein are those of the authors and do not necessarily reflect those of the US Government.

\bibliography{anthology,references/plbart,references/REDCODER}
\bibliographystyle{acl_natbib}

\clearpage
\appendix
\twocolumn[{%
 \centering
 \Large\bf Supplementary Material: Appendices \\ [20pt]
}]
\section{Qualitative Example}
In Figure \ref{qual_example1}, we show an example of generated code by a baseline and different modules of REDCODER. The input summary asks to write a code (in Java) to \highlight{get a MuxerStream given a position}.

We show two of the corresponding retrieved code, their summaries (for \emph{bimodal} instances), generated code of PLBART, \tool, and \toolext. As can be seen, PLBART generates a basic but relevant code; both retrieved code (rank-1 and rank-3) contains the statements with variable \highlight{cPtr} one of them is of \highlight{MuxerStream} class, and another is from \highlight{DeMuxerStream} class. \tool generates a somewhat correct code of \highlight{MuxerStream} class  and it takes the \highlight{position} argument too. Seemingly, while fusing the retrieved code, we suspect that as the tentative function name \highlight{MuxerStream} mentioned in the input summary does not match the function name  \highlight{DeMuxerStream} of the rank-3 retrieved code, it only adapts one line containing \highlight{cPtr} from rank-3 retrieved code (line \#3) and takes the rests including the function definition (i.e., line \#1) from the rank-1 retrieved code. Now when \toolext is allowed to leverage the summaries of the retrieved code, it can match the summary of the rank-3 retrieved code with the input, and that is why it produces the \highlight{MuxerStream} class object but with the \highlight{throw exceptions} from the rank-3 retrieved code.

\section{Performance Difference of PLBART on CodeXGLUE and Concode}
Concode is a relatively easier dataset for code generation and retrieval due to several pre-processing steps taken by its authors. Along with additional contexts (environment variables and methods) in the input summary, Concode artifacts the target code by replacing the specific variable names with generic tokens. 

\lstset{escapechar=@,style=CustomJava}
\begin{lstlisting}[
    basicstyle=\fontsize{10}{11}\selectfont\ttfamily,
]
void function(Element arg0, 
    Formula arg1) {
    arg0.addElement(
        "concode_string").setText(
            arg1.getText());
}
\end{lstlisting}

Therefore, we suspect that due to this, PLBART achieves good EM score for Concode but not for the generation of real code in CodeXGLUE.

Analogously for the retrieval models, code retrieved by BM25 have also a large word overlapping with the targets in Concode in contrast to CodeXGLUE (1st row in Table \ref{table:csnet_gen} and \ref{table:concode}). Consequently, BM25 retrieval boosts PLBART (i.e., BM25 + PLBART) more in Concode than that in CodeXGLUE (3rd row for the bottom in Table \ref{table:csnet_gen} and \ref{table:concode}). Overall, we anticipate all these skewness in model performances are due to the dataset characteristics.

\begin{table*}[!htb]
\centering
\begin{tabular}{l|l|l|c|c|c|c|c}
    \hline
    \multirow{3}{*}{Dataset} &  \multirow{3}{*}{Lang.}  & \multirow{3}{*}{Task} & \multicolumn{3}{c|}{Retrieval Database} & 
    \multirow{3}{*}{|Size|} &
    \multirow{3}{*}{|Nonparallel|}\\
    \cline{4-6}
      &  &  &{CSNet} &   {CCSD} &  {Concode}  &  & \\
      \hline
     \multirow{4}{*}{CodeXGLUE} & \multirow{2}{*}{Python}  & Gen. & \cmark & \xmark & \xmark & 1.2M & 504K\\
     &   & Sum. & \cmark & \cmark & \xmark & 1.1M & 833K \\
     \cline{2-8}
     & \multirow{2}{*}{Java}  & Gen.&  \cmark & \xmark & \xmark & 1.6M & 543K\\
     &   & Sum. & \cmark & \cmark & \xmark & 1.1M & 903K \\
     \hline
    Concode & Java  & Gen. & \xmark & \xmark & \cmark & 104K & 0 \\
    \hline
\end{tabular}
\caption{Retrieval database statistics. ``Size'' refers to both of parallel and nonparallel code or summaries. As Concode has a different data format, we only retrieve from itself. Nonparallel means the retrieval candidates are only code (for code gen.) and only summaries (for code sum.). CSNet (CodeSearchNet), CCSD refer to \citet{husain2019codesearchnet} and  \citet{liu2021retrievalaugmented}.}
\label{tab:ret_db_stat}
\end{table*}

\begin{table*}[!htb]
\centering
\resizebox{\linewidth}{!}{%
\begin{tabular}{l| l| l| c c c| c c c }
\hline
code  & target present & summary & \multicolumn{3}{c|}{CodeXGLUE (Java)} &  \multicolumn{3}{c}{CodeXGLUE (Python)}  \\ 
\cline{4-9}
retrieval & in retrieval & retrieval & BLEU & EM & CodeBLEU & BLEU & EM & CodeBLEU  \\
\hline
\xmark &\xmark & \xmark &  10.1 & 0.0 & 14.96 & 4.89  & 0.0  & 12.01  \\
\hline
\multirow{4}{*}{\cmark} & \multirow{2}{*}{\xmark} & \xmark & {26.92} & {8.95} & {31.15} & { 22.74} & {8.88} & {28.93}  \\
 &  & \cmark & {\bf 28.98} & {\bf 10.21} & {\bf 33.18} & {\bf 24.43} & {\bf 9.61} & {\bf 30.21} \\
\cline{2-9} 
& \multirow{2}{*}{\cmark} & \xmark & 36.33 & {29.41} & {41.38} & {32.14} & {27.48} & {38.02 }  \\
&  &  \cmark &{\bf 42.82} & {\bf 36.99}  & {\bf 47.25} &  {\bf 38.87} & {\bf 34.51} & {\bf 43.78} \\
\hline
\end{tabular}
}
\caption{
Ablation results on source code generation using the retrieved code and its summary together when the reference target code is absent and present in the retrieval database respectively.  
}
\label{table:code_translation}
\end{table*}

\begin{table*}[!htb]
\centering
\begin{tabular}{l|c@{\hskip 0.15in} c | c@{\hskip 0.15in} c }
\hline
\multirow{2}{*}{Methods}  & \multicolumn{2}{c|}{CodeXGLUE-Python} & \multicolumn{2}{c}{CodeXGLUE-Java} \\
\cline{2-5}
 & BLEU-4 & ROUGE-L & BLEU-4 & ROUGE-L \\
\hline
 \coder  & 46.6 & 53.8  & 48.0 & 55.7 \\
\tool& { 47.0} & { 55.4} & {\bf 50.4} & {\bf 58.8} \\
\toolext & {\bf 47.1} & {\bf 55.5} & {\bf 50.4} & { 58.7} \\
\hline
\end{tabular}
\caption{
Evaluation results of  code summarization keeping the target summary in the retrieval database.}
\label{tab:sum_improve_with_ref}
\end{table*}

\begin{figure*}[!htb]
\minipage{0.5\textwidth}
  \includegraphics[width=1.0\linewidth]{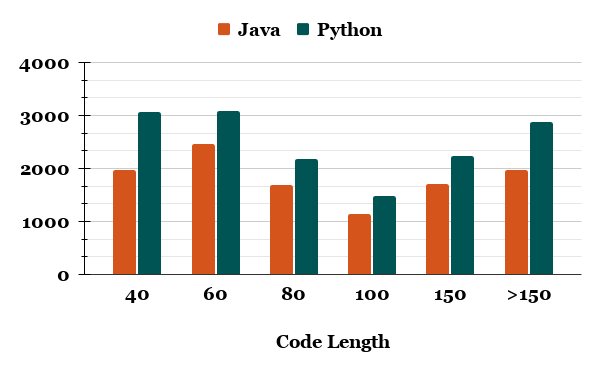}
  \caption{\#Code per target  length.}\label{fig:num_examples_vs_len}
\endminipage
\hfill
\minipage{0.5\textwidth}%
  \includegraphics[width=1.0\linewidth]{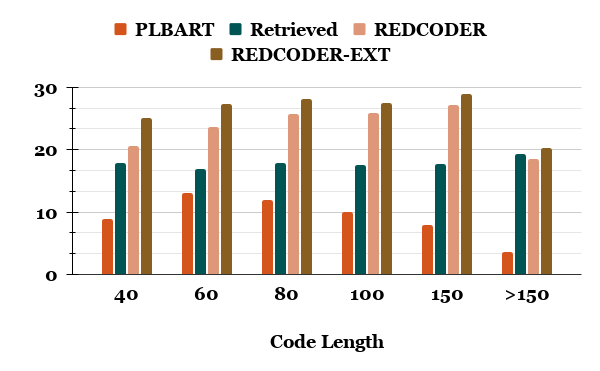}
  \caption{BLEU vs target len. (Java)}\label{fig:bleu_vs_len_java}
\endminipage
\vspace{-30pt}
\end{figure*}


\clearpage

\begin{figure*}[t]
\centering
\rule{\linewidth}{0.9pt}
{\Large{\bf Input summary:} Get the MuxerStream at the given position.}
\\ \vspace{5mm}
\begin{subfigure}{\textwidth}
\captionsetup{font=large}
\caption{\underline{PLBART Prediction} {\bf[BLEU: 0.1439]}}
\begin{tabular}{l}
\lstset{escapechar=@,style=CustomJava}
\begin{lstlisting}[
    basicstyle=\fontsize{10}{11}\selectfont\ttfamily,
]
public MuxerStream getMuxerStream (int position) {
    if (muxerStream == null) {
        muxerStream = new MuxerStream (this, position)
    }
    return muxerStream;
}
\end{lstlisting}
\end{tabular}
\end{subfigure}%
\vspace{5mm}
\begin{subfigure}{\textwidth}
\lstset{escapechar=@,style=CustomJava}
\captionsetup{font=large}
\caption{\underline{SCODER Retrieval (Rank-1)} {\bf[BLEU: 0.4298]}}
\begin{tabular}{l}
\lstset{escapechar=@,style=CustomJava}
\begin{lstlisting}[
    basicstyle=\fontsize{10}{11}\selectfont\ttfamily,
]
public Muxer getMuxer () {
    long cPtr = VideoJNI.MuxerStream_getMuxer(swigCPtr, this);
    return (cPtr == 0) ? null : new Muxer (cPtr, false);
}
\end{lstlisting}
\end{tabular}
\\ \\ 
{{\bf Paired summary:} Get the Muxer that this stream belongs to} 
\\ \vspace{5mm}
\end{subfigure}
\vspace{5mm}
\begin{subfigure}{\textwidth}
\lstset{escapechar=@,style=CustomJava}
\captionsetup{font=large}
\caption{\underline{SCODER Retrieval (Rank-3)} {\bf[BLEU: 0.6740]}}
\begin{tabular}{l}
\lstset{escapechar=@,style=CustomJava}
\begin{lstlisting}[
    basicstyle=\fontsize{10}{11}\selectfont\ttfamily,
]
public DemuxerStream getStream (int streamIndex) 
    throws java.lang.InterruptedException, java.io.IOException {
    long cPtr = VideoJNI.Demuxer_getStream(swigCPtr, this, streamIndex);
    return (cPtr == 0) ? null : new DemuxerStream (cPtr, false);
}
\end{lstlisting}
\end{tabular}
\end{subfigure}
\vspace{5mm}
\begin{subfigure}{\textwidth}
\lstset{escapechar=@,style=CustomJava}
\captionsetup{font=large}
\caption{\underline{REDCODER Prediction} {\bf[BLEU: 0.6170]}}
\begin{tabular}{l}
\lstset{escapechar=@,style=CustomJava}
\begin{lstlisting}[
    basicstyle=\fontsize{10}{11}\selectfont\ttfamily,
]
public MuxerStream getMuxerStream (int position) {
    long cPtr = VideoJNI.MuxerStream_getMuxer(swigCPtr, this, position);
    return (cPtr == 0) ? null : new MuxerStream (cPtr, false);
}
\end{lstlisting}
\end{tabular}
\end{subfigure}
\vspace{5mm}
\begin{subfigure}{\textwidth}
\lstset{escapechar=@,style=CustomJava}
\captionsetup{font=large}
\caption{\underline{REDCODER-ext Prediction} {\bf[BLEU: 0.8062]}}
\begin{tabular}{l}
\lstset{escapechar=@,style=CustomJava}
\begin{lstlisting}[
    basicstyle=\fontsize{10}{11}\selectfont\ttfamily,
]
public MuxerStream getMuxer (int streamIndex) 
    throws java.lang.InterruptedException, java.io.IOException {
    long cPtr = VideoJNI.MuxerStream_getMuxer(swigCPtr, this, streamIndex);
    return (cPtr == 0) ? null : new MuxerStream (cPtr, false);
}
\end{lstlisting}
\end{tabular}
\end{subfigure}
\vspace{5mm}
\begin{subfigure}{\textwidth}
\lstset{escapechar=@,style=CustomJava}
\captionsetup{font=large}
\caption{\underline{Reference (Gold Output)}}
\begin{tabular}{l}
\lstset{escapechar=@,style=CustomJava}
\begin{lstlisting}[
    basicstyle=\fontsize{10}{11}\selectfont\ttfamily,
]
public MuxerStream getMuxer (int streamIndex) 
    throws java.lang.InterruptedException, java.io.IOException {
    long cPtr = VideoJNI.MuxerStream_getMuxer(swigCPtr, this, streamIndex);
    return (cPtr == 0) ? null : new MuxerStream (cPtr, false);
}
\end{lstlisting}
\end{tabular}
\end{subfigure}
\rule{\linewidth}{0.9pt}
\caption{
A qualitative example to show the effectiveness of retrieval-augmented generation as proposed in REDCODER framework
}
\label{qual_example1}
\end{figure*}


\end{document}